# Time-Resolved Ferromagnetic Resonance in Epitaxial Fe$_{1-x}$Co$_x$ Films

D. M. Engebretson,[1] S. Zellinger,[1] L. C. Chen,[2,*] C. J. Palmstrøm,[2] and P. A. Crowell[1]

[1]School of Physics and Astronomy, University of Minnesota, Minneapolis, MN 55455

[2]Department of Chemical Engineering and Materials Science, University of Minnesota, Minneapolis, MN 55455

Magnetodynamics in epitaxial Fe$_{1-x}$Co$_x$ films on GaAs (100) are studied using time-resolved ferromagnetic resonance, in which the free precession of the magnetization after an impulsive excitation is measured using the polar Kerr effect. The sample is rotated with respect to the static and pulsed field directions, providing a complete mapping of the free energy surface and characteristic relaxation times. The magnetic response can be simulated with a simple coherent rotation model except in the immediate vicinity of switching fields. Bulk and surface anisotropies are identified, and unusual dynamics associated with the coexistence of cubic and uniaxial anisotropies are observed.

An interest in the fundamental limits facing magnetic recording technologies has motivated a closer examination of the mechanisms through which switching and magnetic reversal occur. For example, it is now possible to probe the effect of impulsive excitations with bandwidths that exceed typical ferromagnetic resonance frequencies.[1] The effective energy barriers for reversal in this regime are determined by the magnetocrystalline and shape anisotropies. For sufficiently large pulses, the switching speed is limited only by the precession frequency and the lifetime of excitations generated by the pulse. The recent development of real-time optical and electronic techniques for studying magnetization dynamics has provided new tools for studying high-frequency behavior. Free precession of the magnetization has been observed on picosecond time scales in permalloy and iron,[2-5] and complete reversal has been observed in permalloy on nanosecond time scales.[6,7] This article presents time-resolved resonance measurements made on single-crystal ferromagnetic films. The samples are thin films of $Fe_{1-x}Co_x$ grown on GaAs (100) by molecular beam epitaxy.[8] A high-bandwidth (~ 10 GHz) magnetic field pulse with a variable orientation is used to tip the sample magnetization, which is then probed using the polar Kerr effect. We demonstrate that the time-domain response can be used to map out the free energy surface of the system. Rich dynamical behavior is observed due to the superposition of both bulk and surface-induced anisotropies.

The magnetic properties of epitaxial $Fe_{1-x}Co_x$ grown on (2×4)c(2×8) reconstructed GaAs (100) are determined by the combined effects of a bulk cubic anisotropy and an extrinsic uniaxial anisotropy that is due to the two-fold symmetry of the GaAs (100) surface.[8] The magnitudes of the two anisotropies are comparable for films with thicknesses of the order of 200 Å and Co compositions $x \approx 0.5$. Typical hysteresis loops for in-plane fields are shown in Fig. 1 for a sample with a nominal thickness of 160 Å and Co composition $x = 0.51$. The salient feature of these data is the jump in the magnetization at approximately 130 Oe for fields applied along the $[01\bar{1}]$ direction. This is a direct consequence of the fact that $[01\bar{1}]$ is a hard direction for the uniaxial anisotropy and an easy axis of the four-fold cubic anisotropy. The hysteretic behavior in Fig. 1 can be understood with a simple coherent rotation model using the free energy

$$F = -\mathbf{M} \cdot \mathbf{H} + \frac{K_v}{4} \sin^2 2\varphi + K_u \cos^2(\varphi - \frac{\pi}{4}), \qquad (1)$$

where $K_v$ and $K_u$ are the cubic and uniaxial anisotropies and $\phi$ is the azimuthal angle measured from [010]. (The demagnetizing energy is sufficiently large so that the magnetization lies in the plane.) The total magnetocrystalline anisotropy has deep local minima along [011]. For fields along $[01\bar{1}]$, this competes with the Zeeman energy, and there is a nearly 90 degree jump in the direction of the magnetization when the barrier between the minima disappears. For the sample considered here, this occurs at $H_s \sim 130$ Oe, and the cubic and uniaxial anisotropies are $K_v = -2.2 \times 10^5$ erg/cm$^3$ and $K_u = -2.4 \times 10^5$ erg/cm$^3$. This model has been confirmed on a variety of samples using the transverse and longitudinal Kerr effects as well as vibrating sample magnetometry.[9]

This article focuses on the magneto-dynamics of these films under pulsed excitation. As in recent work by Freeman and co-workers on polycrystalline films,[2, 5] we use a time-resolved technique in which a magnetic field pulse is generated by terminating a fast photodiode (rise time ~ 60 psec) in a 50 ohm stripline, tapered to a width of approximately 30 microns. The photodiode is pumped by 150 fsec pulses from a Ti:sapphire laser, and the z-component of the magnetization is probed by measuring the Kerr rotation of a time-delayed probe pulse. The pump pulse is modulated by an optical chopper, and the Kerr signal is measured by detecting the off-balance signal from a polarization bridge using a lock-in amplifier referenced to the chopper frequency. The $Fe_{1-x}Co_x$ films are patterned into disks of diameter 20 µm by photolithography and wet etching, and the substrates are thinned to approximately 25 µm by mechanical and chemical polishing. The sample is then placed on the stripline, on which it can be rotated. For the measurements reported here, the angle between the pulsed and static fields remains fixed at 90 degrees as shown in the schematic diagram of Fig. 2. The ease of rotation and the ability to work with a free-standing crystalline sample are the significant advantages of our approach, although it entails some degradation in pulse amplitude and uniformity.

Time-domain data are shown in Fig. 3 for magnetic fields of approximately 150 Oe applied along the [011] (easy), [010] (intermediate) and [01$\bar{1}$] (hard) directions. The sample is the same as used for the measurements of Fig. 1. The easy and intermediate axes show single precession frequencies of 10.2 ± 0.1 GHz. and 9.4 ± 0.10 GHz as determined from fast Fourier transforms (FFT's) of the time-domain data. The data for the applied field along the [01$\bar{1}$] direction are qualitatively different, showing the apparent mixing of two frequencies (1.8 GHz and 7.2 GHz). These data were obtained for fields just above the jump field $H_s$ ~ 130 Oe indicated in Fig. 1. This effect is much larger than the weak beating at short times that is observed along the other directions, which we attribute to non-uniform excitation by the microwave field.

The precession frequencies determined from FFT's are shown in Fig. 4 as a function of DC field applied along the [011], [010] and [01$\bar{1}$] directions. For fields along [011], the precession frequency increases monotonically with field, while it remains nearly flat for fields up to 600 Oe applied along [010]. The evolution with field along [01$\bar{1}$] is qualitatively very different, showing a pronounced drop in the vicinity of the jump field $H_s$. On the low-field side of the jump, the static magnetization is close to the [011] direction, while it is along [01$\bar{1}$] above 150 Oe. Qualitatively, this behavior is similar to that observed in an ideal uniaxial ferromagnet,[10] in which the resonance frequency drops to zero when the applied field perpendicular to the easy axis is equal to the anisotropy field and then rises again as the field is increased further. In the system studied here, the addition of a cubic anisotropy leads to a first-order jump instead of continuous rotation.

A quantitative interpretation of the data shown in Fig. 4 follows from a coherent rotation model including Zeeman, anisotropy, and demagnetizing energies. The most general form of the free energy is

$$F = -\mathbf{M} \cdot \mathbf{H} + \frac{K_v}{4}(\sin^2 2\theta + \sin^2 2\varphi \sin^4 \theta) \\ + K_u \sin^2 \theta \cos^2(\varphi - \frac{\pi}{4}) + (2\pi M^2 + K_\perp)\cos^2 \theta, \quad (2)$$

where θ and φ are the polar and azimuthal angles measured from [100] and [010] respectively. We fix $M = 1900$ emu/cm$^3$, $K_v = -2.2 \times 10^5$ erg/cm$^3$, and $K_u = -2.4 \times 10^5$ erg/cm$^3$ as determined from a complete set of static in-plane magnetization measurements as a function of field orientation, leaving the perpendicular anisotropy $K_\perp$ as the only free parameter. The set of solid curves shown in Fig. 4 for the three different field orientations represents a best fit of the precession frequencies with $K_\perp = (-5.0 \pm 0.5) \times$ $10^6$ erg/cm$^3$. This is an exceptionally large value of $K_\perp$ for a film with a thickness of 160 Å. However, it is in agreement with the anisotropy field determined from a polar Kerr effect measurement, and a similarly large value of $K_\perp$ was found in cavity FMR experiments on similar samples.[11] Although tetragonal distortions have been shown to produce significant surface anisotropies in thinner films,[12] it is possible that some of the discrepancy in this case could be due to an error in the film thickness and hence the magnetization. A systematic measurement of $K_\perp$ as a function of film thickness would be required to address this question.

Although the coherent rotation model provides a good fit of the observed precession frequencies, it does not provide an adequate description of the dynamics observed for fields along the [01$\bar{1}$] direction. Within the coherent rotation picture, we have attempted to model the time response with the Landau-Lifshitz-Gilbert equation,

$$\frac{\partial M}{\partial t} = -\gamma\left(M \times H_{eff}\right) + \frac{\alpha}{M_s}\left(M \times \frac{\partial M}{\partial t}\right), \quad (3)$$

where the effective field $H_{eff} = -\nabla_M F$, $M_s$ is the saturation magnetization, and $\alpha$ is the phenomenological damping parameter. We use the free energy $F$ defined in Eq. 2. For fields along the easy direction, we obtain reasonable qualitative agreement with damping parameters $\alpha$ in the range 0.004 – 0.012 for applied fields from 0 Oe to 600 Oe. The smallest damping was observed at a field of 325 Oe. In spite of the erratic evolution of the damping parameter, which may be due to edge domains, the easy-axis behavior is consistent with nearly homogeneous precession. Along the [01$\bar{1}$] direction, however, the model shows neither the multiple frequencies nor the long time-scale signatures that appear in the data of Fig. 3(c). Given the coexistence of two minima along [011] and $[01\bar{1}]$, one would expect two frequencies to appear due to domain structure in the hysteresis region around the jump field, but it is not evident why the low-frequency response should be so long-lived. A detailed understanding of this behavior will require a full micromagnetic treatment of the experimental system.

This work was supported by the Research Corporation, NSF DMR-9983777, the Alfred P. Sloan Foundation, and ONR grant N/N00014-99-1-0233 (CJP).

**Figure Captions**

FIG. 1. Magnetic hysteresis loops for applied fields along [011] (easy axis), [010] and [01$\bar{1}$]. Note the appearance of the secondary loops along [01$\bar{1}$]. The split field $H_s$ is indicated by the vertical dashed line.

FIG. 2. A schematic diagram of the experiment, showing the orientation of the pulsed ($H_1$) and static ($H_0$) fields. The sample substrate, which is approximately 25 μm thick, can be rotated in the plane of the figure as shown.

FIG. 3. The time evolution of magnetization is plotted for fields along the three principal directions.

FIG. 4. The resonant frequencies (symbols) determined from Fourier transforms of the time-domain data are shown as a function of the magnetic field applied along the three principal directions. The solid curves are fits to the coherent rotation model described in the text.

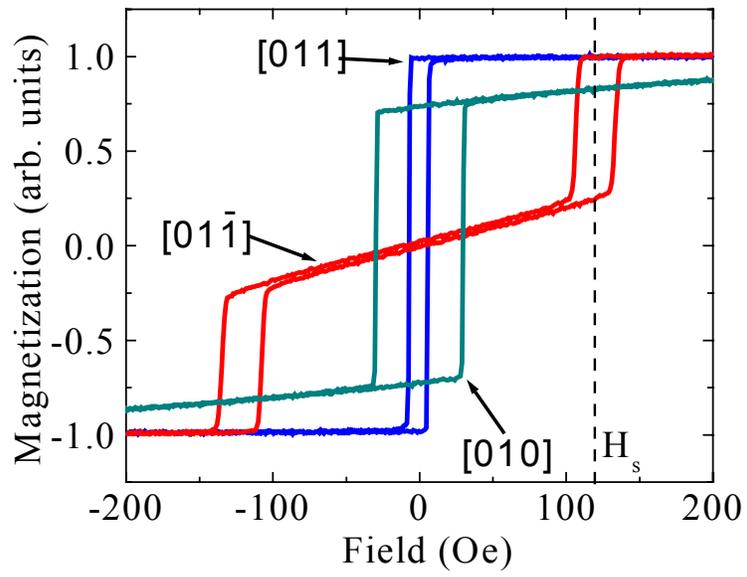

FIG. 1. Engebretson *et al.*

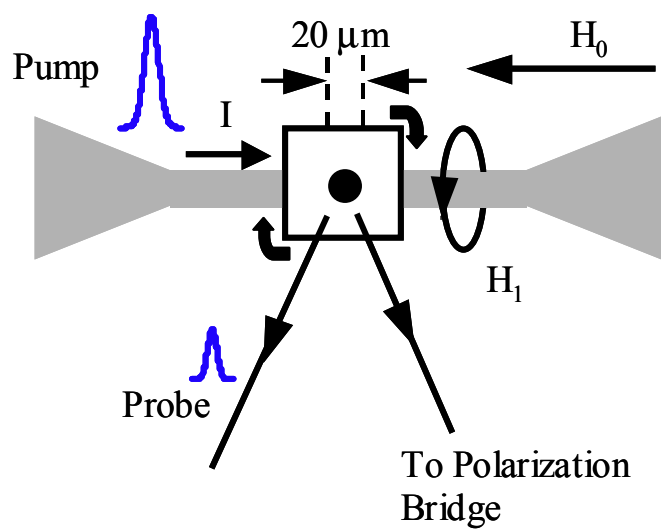

FIG. 2. Engebretson *et al.*

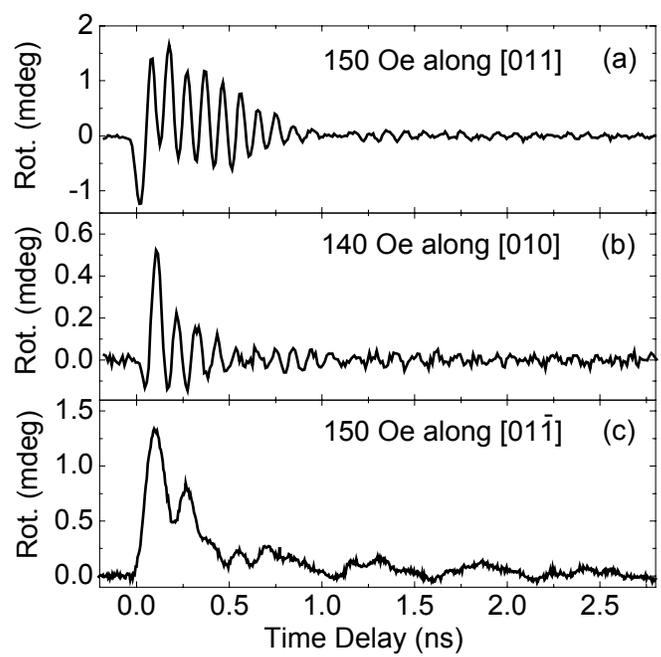

FIG. 3. Engebretson *et al.*

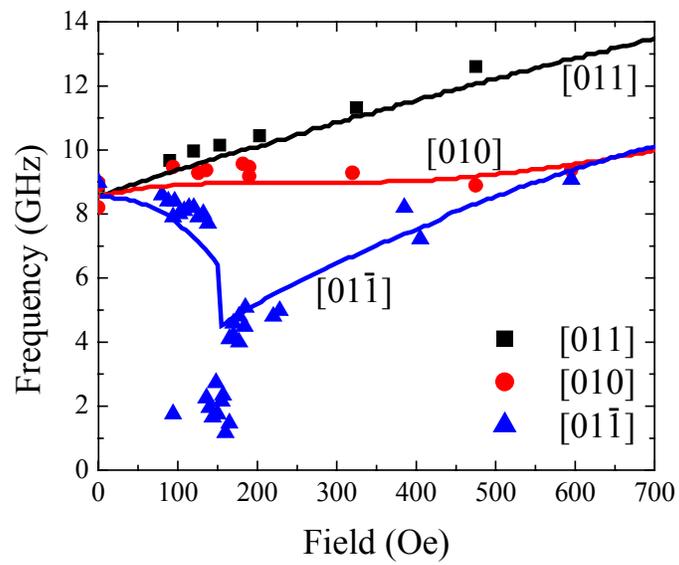

FIG. 4. Engebretson *et al.*